\documentclass[twocolumn,pre,showpacs]{revtex4}

\usepackage{graphicx}
\usepackage{rotating}
\usepackage{amsmath}
\usepackage{amsfonts}
\usepackage{amssymb}
\usepackage{enumerate}
\usepackage{longtable}
\usepackage{dcolumn}
\usepackage{bm}
\usepackage{epsfig}
\usepackage[colorlinks]{hyperref}
\usepackage{color}
\begin{document}

\title{Social structure formation in a network of agents playing a hybrid of ultimatum and dictator games}
\author{Jan E. Snellman$^{1}$}
\author{Rafael A. Barrio$^{2}$}
\author{Kimmo K. Kaski$^{1,3}$}
\affiliation{$^1$Department of Computer Science, Aalto University School of Science, FI-00076 AALTO, Finland}
\affiliation{$^2$Instituto de F{\'i}sica, Universidad Nacional Aut{\'o}noma de  M{\'e}xico, 01000 CDMX, Mexico}
\affiliation{$^3$The Alan Turing Institute, 96 Euston Rd, Kings Cross, London NW1 2DB, UK}
\date{\textrm{\today}}

\begin{abstract}

Here we present an agent-based model where agents interact with other agents by playing a hybrid of dictator and ultimatum games in a co-evolving social network. The basic assumption about the behaviour of the agents in both games is that they try to attain superior socioeconomic positions relative to other agents. As the model parameters we have chosen the relative proportions of the dictator and ultimatum game strategies being played between a pair of agents in a single social transaction and a parameter depicting the living costs of the agents. The motivation of the study is to examine how different types of social interactions affect the formation of social structures and networks, when the agents have a tendency to maximize their socioeconomic standing. We find that such social networks of agents invariably undergo a community formation process from simple chain-like structure to more complex networks as the living cost parameter is increased. The point where this occurs, depends also on the relative proportion of the dictator and ultimatum games being played. We find that it is harder for complex social structures to form when the dictator game strategy in social transactions of agents becomes more dominant over that of the ultimatum game.
\end{abstract}
\maketitle

\section{Introduction}
\label{intro}

Studies of human social behaviour suggests that striving for superior social positions is a fundamental characteristic of human nature \cite{AHH2015}, and this feature has even measurable effects on human brain activity \cite{ZTCB2008,ISS2008,KMD2012,UP2014}. That being the case, however, one could ask why the resultant competition between individuals does apparently not preclude humans from forming structured societies? This is an intriguing question since humans are inherently sociable and show clear tendency to form complex social and societal structures. An obvious answer to this question is that humans are dependent on each other for their survival and well-being, which in turn implies that without either environmental or social pressure humans would not develop complex communities at all (see, e.g. \cite{ASGB2014} and references within). 

In order to model the societal effects of this tendency of superiority maximisation we introduced the better-than-hypothesis, or BTH, in \cite{SGGBK2017}. As the name implies, the working assumption of BTH is that humans are motivated primarily by the aim to maximize their social status in society, an assumption which is shared by Alfred Adler's school of individual psychology \cite{adler}. In \cite{SGKBK2018} we discovered an abrupt behavioural change in a model derived from BTH, which could be interpreted as the formation of social structures due to outside pressure. 

Human communities are by nature connected and perpetually changing, so they can be considered as temporal  social networks of individuals. Consequently there has been a number of attempts at the theoretical level to model the community formation process using network theory. There are naturally many different aspects related to the subject in addition to the influence of outside pressures, and thus most studies have concentrated so far on such aspects as the roles of opinion formation \cite{HN2006,IKKB2009,SGGBK2017,LGAGI2005,JA2005,GK2006,SM2013}, deception \cite{IGDKB2014,BGDIK2015} or information exchange \cite{EZ2000,W2015}. 

The model we studied in \cite{SGKBK2018} concerned a social network with agents playing dictator game with their neighbours. The agents were allowed to change their own connections, and the model was governed by two parameters, one of which, the memory parameter measuring how fast the agents forget the way they were treated, and the other, the cost parameter measuring, the proportion of money spent on living costs. We found that when one varies either one of these parameters the social networks produced by the model consist of disconnected chains that lengthen until the parameters reach certain values, after which the networks become much more connected and complex. Such social structure formation process was clearly visible in the shortest path length and susceptibility measures plotted as functions of these parameters, in terms of exhibiting sharp peaks at certain points. 

The process of community formation has been studied previously in models of social networks, observing a structural phase change taking place, see e.g. \cite{RB2002,DM2002,N2003,SPS2012a,SPS2012b,SP2013,EMV2006,HN2006,MJTKK2019}. In the network context phase changes are of particular interest because of the insights they provide into various social phenomena such as community, opinion, and hierarchy formation. In \cite{HN2006}, for instance, the combined effect of co-evolving opinions and social network structure on community and opinion formation was studied, with a phase transition taking place as the relative influence of the two processes was varied. Similar, but more general, approaches were used in \cite{EMV2006} and \cite{MJTKK2019} to study the role of homophily in the dynamics of the social networks and in the segregation of network structures, respectively. On the other hand in \cite{CCV2016} hierarchy formation was presented as a result of  community formation when social and cognitive constraints were involved.

The original dictator game mentioned above is a problem in decision theory first proposed in \cite{KKT1986}, where one of the players, the dictator, is tasked with dividing a fixed reward between the players at will, with no input from the other player, who will get only what the dictator allocates to them. It is different from the closely related ultimatum game \cite{GT1990}, in which the other player, called accepter in the context of the game, is allowed to either reject or accept the offer. In case of rejection, neither player gets anything, while in the case of acceptance the division is realised as it stands. Both games were constructed to demonstrate the limitations of rational economic behaviour. 

The ultimatum and dictator games allow the testing of human economic behaviour, especially the assumption of rationality (see e.g.\cite{S1955} and \cite{HS}). Providing that humans rationally seek to maximize their own profits, in the dictator game the dictators should reserve everything for themselves, while in the ultimatum game one could expect that the accepter should accept any nonzero amount offered, and that the proposers should offer the minimum possible amounts. However, in the actual experiments with real humans the proposers in the ultimatum game and dictators of the dictator game have a tendency to offer substantial amounts to the other player, and unbalanced propositions in the ultimatum game (especially those in favour of the proposer) tend to be rejected by the accepters (see, for example, \cite{E2011,metalyysi,CC2008,FLK2015}). These results immediately raise the question on the rationality assumption of human behaviour. As we argued in \cite{SGGBK2017}, assuming humans to rationally seek superior position over the others means in the case of the ultimatum game that accepters should accept offers of over $1/3$ of the full amount, and that the proposers should realise this and adjust their offers accordingly\footnote[1]{For the sake of completeness it is of interest to briefly repeat the derivation of this result here. If an accepter is offered proportion $a$ of a total amount of $M$ in a one shot standard ultimatum game, the BTH utility function (Eq. \ref{uprime} presented below) becomes $a + (a - (M - a)) = 3a - M$. The accepter will only accept if this is positive, i.e. $a > M/3$. It should be noted that if one were to use different weights for absolute and relative terms of Eq. \ref{uprime}, the result would depend on the weights.}. This analytical result seems to be in agreement with experimental data of real people playing the game.

In this study we extend our earlier model presented in \cite{SGKBK2018} by letting the agents play ultimatum game as well as dictator game, and take a look at the effects this has on the observed social structure and network formation. The motivation for this study is to test, how different modes of social interactions change the structures and dynamics of social networks of humans following the better than hypothesis (BTH). In this new model, the agents choose at random the games they play according to their internalised social norms, which are represented simply by a parameter $d$ describing the probability of the dictator game being chosen over the ultimatum game. A special area of interest is to study the interplay between {\em unilateralism} in social interactions represented by the dictator game and a limited {\em reciprocity} represented by the ultimatum game. 

This paper is organised such that in the next Section \ref{themodel} we define the network model of agents playing the game with a strategy that is a hybrid of the ultimatum and dictator game strategies. This is followed by the simulation results presented in Section \ref{results}. Finally we draw conclusions in Section \ref{con}.

\section{Network model of agents playing ultimatum and dictator games}
\label{themodel}

Let us consider a set of linked agents forming a social network and playing a hybrid of ultimatum and dictator games with all their network neighbours. In a single step of the game step each agent chooses which other to play with, and how generous to be with those agents. The former choice constitutes network rewiring and the latter changing generosity. In what follows we will illustrate how each one of these choices result in and how they relate to each other. Let us start by describing the  hybrid game. 

In both the ultimatum and dictator game each agent has its individual offering rate $x_i$, but in the case of the ultimatum game each agent has in addition a threshold $y_i$ for accepting the offer (see \cite{X2010}). In our hybrid model we assume that each agent has a choice between two different social strategies, playing either the dictator game or the ultimatum game with their neighbors, between which they make the choice at random. The simulation proceeds in cycles, in each of which the agents play either the dictator game with probability $d$ or the ultimatum game with probability $(1 - d)$, with all the agents they are connected to. As stated in the introduction, parameter $d$ is interpreted to represent internalised social norms of the agents that determine, which game is appropriate in which social circumstances. The game being played is chosen independently for each transaction between the agents, so agent $i$ could play ultimatum game with agent $j$ and dictator game with agent $k$ in a single cycle. Also, since all the agents get to play the game in the role of proposer/dictator once during the game cycle, all the pairs play twice in a cycle in inverted roles.  The nth cycle of the game is denoted by $T_n$ and the agents play it in the same order. This is based on the assumption that playing the game in either fixed or random order would not make much difference. As explained below, the modeled agents only update their behavioural patterns in the game in between the rounds of play, and, additionally, the changing wealth or any other circumstances of the agents during the play rounds do not have effect on their behaviour in the game. In order to test this assumption we will make a few tests with random playing order, as explained in more detail in Sec. \ref{results}

The hybrid double strategy model differs from the single strategy model of Ref. \cite{SGKBK2018} such that with probability $(1 - d)$, when the ultimatum game is played, the agent receiving the offer gets the choice of either accepting or rejecting the division of funds offered by the proposing agent. Then if agent $i$ makes an offer to agent $j$ the transaction occurs if either the dictator game is played or in case of the ultimatum game if $x_i \ge y_j$. In the case the division of funds is realised the accumulated wealth $v_i$ and $v_j$ of the agents $i$ and $j$, respectively, change as follows: 
\begin{eqnarray}
v_i(t_1) &=& v_i(t_0) + (1 - x_i) M, \\
v_j(t_1) &=& v_j(t_0) + x_i M,
\end{eqnarray}
where $M$ is the amount of funds to be divided in each transaction and $t_0$ is the time before the transaction and $t_1$ is the time immediately after it. 
In each cycle of the hybrid game, the amount $cM$, denoting the "living costs", is deducted from the accumulated wealth of the agents, such that the wealth of an agent can only be reduced to zero. Here the cost parameter $c$ turns out to be very important in shaping the structure of the network when the game is played. Different values of the relative proportions $d$ and $(1 - d)$ of the dictator and ultimatum games played by the agents correspond to different rules of social interaction while the parameter $c$ can be interpreted as an external pressure or need forcing the agents to cooperate. 

Now, the wealth acquired by %the 
agent $i$ in a cycle can be written as follows,

\begin{eqnarray}
v_i(T_{n + 1}) =  \max\Big(v_i(T_n) 
&+& M\big( (1 - x_i) | \widehat{\chi}_{ji} | \nonumber \\
&+& \sum_{j \in \widehat{\chi}_{ij}} x_j - c \big) , 0 \Big),
\end{eqnarray}

where $\widehat{\chi}_{ij}$ is the set of those neighbours of agent $i$ that will accept or are forced to accept (depending on whether agent $i$ gets to play either the ultimatum or dictator game with agent $j$) the offer $x_i$, and, conversely, $\widehat{\chi}_{ji}$ is the set of the neighbouring agents' offers $x_j$ the agent $i$ is either willing or forced to accept. If the full set of agents is denoted by $I$ and the set of neighbours of agent $i$ denoted by $m_i$, $\widehat{\chi}_{ij}$ can be written as follows
\begin{equation}
\widehat{\chi}_{ij} = m_i \cap \chi_{ij},
\end{equation}
where $\chi_{ij}$ is the set of all agents that either accept or are forced to accept the offer $x_i$, i.e. 
\begin{equation}
\chi_{ij} = \{ j \in I \vert x_i \geq y_j \lor D_{ij}(T_n) \leq d \},
\end{equation}
where $D_{ij}(T_n)$ is a random variable that determines, whether agent $i$ gets to play the dictator or the ultimatum game with agent $j$. 

Next, we describe the way agents choose their playing partners by cutting or forming social links between each other. This boils down to how their perceived relative status is impacted in their social interactions. According to the BTH (Better Than Hypothesis), agents compare themselves to other agents on the basis of their accumulated wealth. Assuming that agents have knowledge of the wealths of their neighbours and the payoffs they receive from their neighbours mutual neighbours, we can define the utility function of agent $i$ at cycle $T_n$ as follows
\begin{equation}
U_i(T_{n}) =  v_i(T_{n}) + \sum_{l \in m_i} (v_i(T_{n}) - v_l(T_{n})),
\label{deq}
\end{equation}
The terms on the right hand side of Eq. \ref{deq} reflect the status of agent $i$ in absolute (first) and relative (second) terms. In principle these terms could be weighted differently using purpose-built coefficients, but in this work we assume these terms to be equally important for the sake of simplicity. 
Then, we can determine the contribution by agent $j$ to the utility of agent $i$, and write it in the following cumulative form
\begin{equation}
U_{ij}(T_{n+1}) = U_{ij}(T_n) + a_{ij} U'_{ij}(T_{n+1}),
\label{uijeq}
\end{equation}
where $a_{ij}$ is an element of the adjacency matrix of the the network of agents:
\begin{equation}
a_{ij} = 
\left\{
\begin{aligned}
&1,\;\; \mathrm{if}\;\mathrm{agents}\;\mathrm{i}\;\mathrm{and}\; \mathrm{j}\; \mathrm{are}\;\mathrm{linked}\;\;\\
&0,\;\; \mathrm{otherwise},
\end{aligned}
\right.
\end{equation}
and

\begin{eqnarray}
U'_{ij}(T_{n+1}) = 
\left\{
\begin{aligned}
& U^{h}_{ij},\;\; \mathrm{if}\;\; x_j \geq y_i \lor D_{ij}(T_n) \leq d,\\
&-  (v_j(T_{n+1}) - v_j(T_n)),\;\; \mathrm{otherwise}\\
\end{aligned}
\right.
\label{uprime}
\end{eqnarray}
where $n_J$ is the number of neighbouring agents that agents $i$ and $j$ have in common, i.e. agents in $J = m_i \cap m_j$, and
\begin{equation}
U^{h}_{ij} = (k_i(T_n) -n_J + 1)x_j(T_n) M -  (v_j(T_{n+1}) - v_j(T_n)).
\end{equation}

\begin{figure}[ht]
\begin{center}
    \includegraphics[width=0.65\columnwidth]{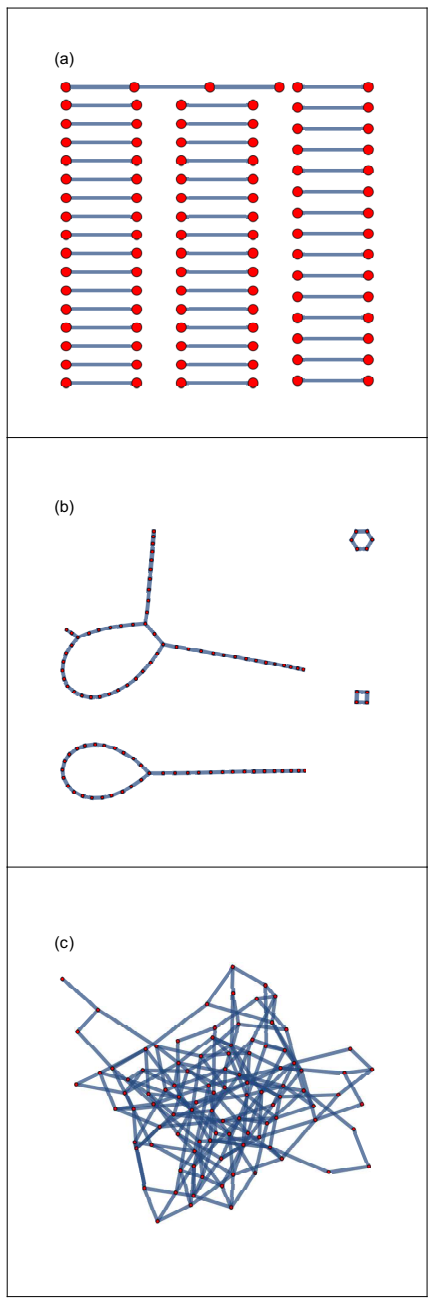}
\caption{The structure of co-evolving social network of $100$ agents interacting socially by playing at random and at the same rate ($d=0.5$) either the dictator or ultimatum games. The three panels show a snapshot of the structure after 10000 time steps for cost parameter values $c = 0.3$ (panel (a)), and $c = 1.1$ (panel (b)) and $c = 3.0$ (panel(c)).}
\label{netFig1}
\end{center}
\end{figure}

\begin{figure*}[ht]
\begin{center}
\includegraphics[width=2.0\columnwidth]{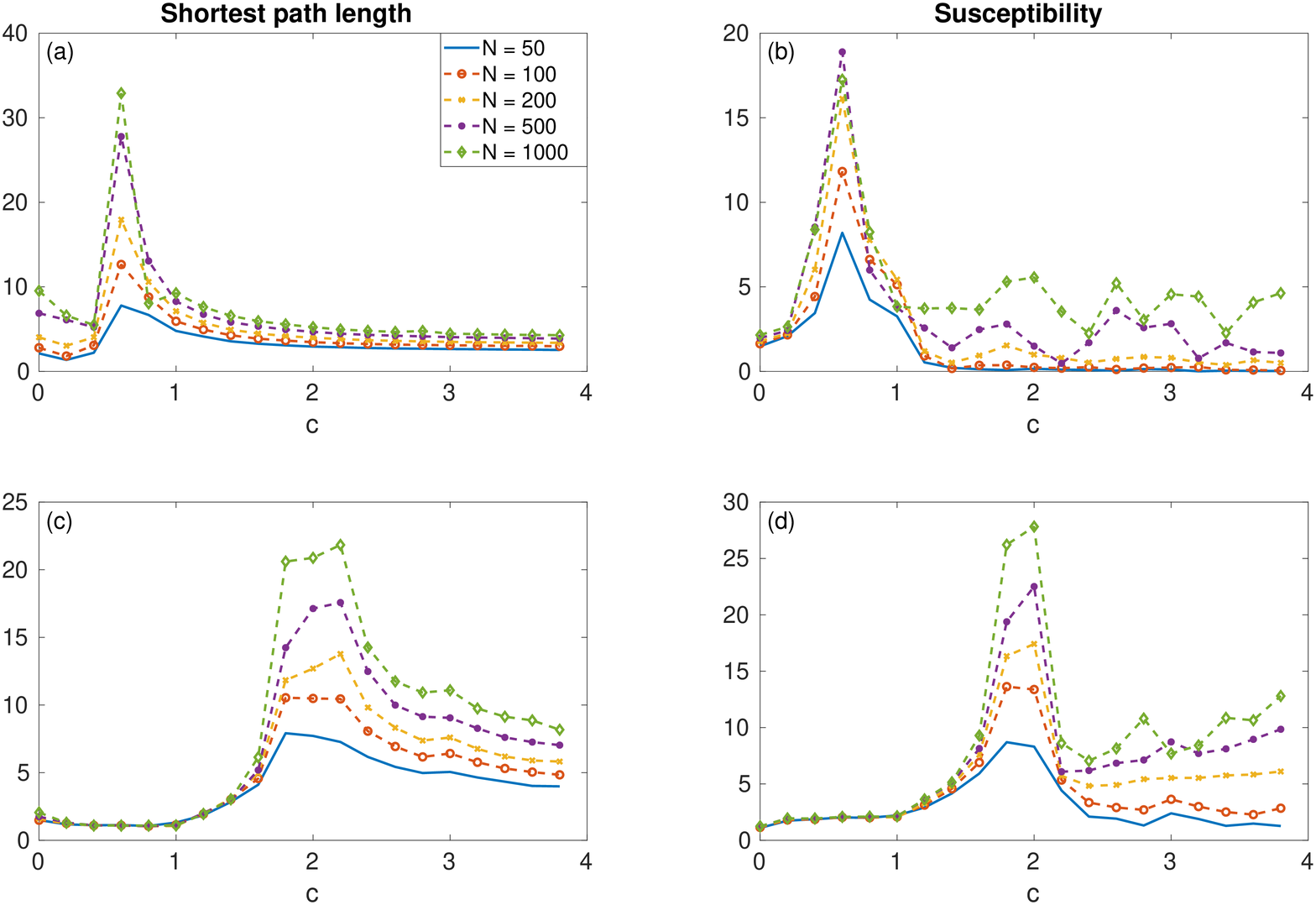}
\caption{The shortest path length and the susceptibility in the pure ultimatum ($d=0$; (a) and (b)) and dictator ($d=1$; (c) and (d)) modes as function of the cost parameter $c$, for for four different network sizes with total number of agents $N$ shown in the legend.}
\label{phasefig}
\end{center}
\end{figure*}

After each cycle agents redefine their social relations based on the contributions of their neighbours to their current utilities, as defined by Eq. \ref{uijeq}. Let us consider two agents, $i$ and $j$, who have contributed to each others utility by $U_{ij}$ and $U_{ji}$, respectively. Now, agent $i$ will form a social link with the agent $j$ if both utilities $U_{ij}$ and $U_{ji}$ are non-negative, provided that the link between them does not already exist. Similarly, the agent $i$ will cut an existing link with agent $j$, if $U_{ij} \le 0$. This dynamics together with the transaction dynamics constitutes the co-evolution of the social network structure.

The final ingredient of the model is that the agents also modify their generosity in between play cycles. This happens in the following way. The agents choose their offering rate $x_i$ and the acceptance threshold $y_i$ according to Eq. (\ref{deq}) using a mix of simple hill-climbing and random walk methods. The agents have the values of $x_i$ and $y_i$ assigned randomly to them at the start of the simulations, and they seek to find better values by choosing a random direction on the $(x_i,y_i)$-plane and changing their $x_i$ and $y_i$ values to that direction as long as their utility (\ref{deq}) is growing, i.e. $\Delta U_i(T_n) = U_i(T_n) - U_i(T_{n-1})$ remains positive. When $\Delta U_i$ becomes negative, the agents choose randomly a new direction to proceed. The whole process can then be written in the following form:
\begin{eqnarray}
x_i(T_{n + 1}) &=&  x_i(T_{n}) + \delta s(T_n)\cos(\phi_i) \nonumber \\
y_i(T_{n + 1}) &=&  y_i(T_{n}) + \delta s(T_n)\sin(\phi_i),
\label{xiyi}
\end{eqnarray}
where $\delta s(T)$ is the length of the step the agents take into their chosen direction $\phi_i$. As stated, the direction $\phi_i$ is the angle that is reassigned a different value every time $\Delta U_i$ is negative, i.e.
\begin{equation}
\phi_i(T_{n + 1}) = 
\left\{
\begin{aligned}
&\phi_i(T_{n}),\;\; \mathrm{if}\;\Delta_i \geq 0\;\;\\
&\phi_r,\;\; \mathrm{otherwise},
\end{aligned}
\right.
\end{equation}
where $0 \le \phi_r \le 2\pi$ is chosen randomly. In the pure dictator game mode $(d = 1)$ the acceptance threshold $y_i$ is not used, and the direction of change of the offering rate $x_i$ is flipped every time $\Delta U_i$ is negative. The initial directions of change of the offering rate can be either increasing or decreasing, and offers are made to all the agents at random. In determining the step length $\delta s(T)$ we have adopted the central aspect of simulated annealing techniques, reducing it linearly from $0.11$ to $0.01$ in the course of the first $1000$ time steps. The idea here is that the agents first try to find their offering and acceptance preferences in a very rough manner, before gradually starting a more fine-tuned search. 

In the next section we present the results of comprehensive computer simulations of the model of networked agents interacting socially with their linked neighbors by playing a hybrid of ultimatum and dictator games. Our main focus in these simulations is to get insight into the co-evolving social network structure and community formation behavior therein. 

\section{Results}
\label{results}

\begin{figure*}[ht]
\begin{center}
\epsfxsize = 2.0\columnwidth \epsfysize = 0.65\textheight \epsffile{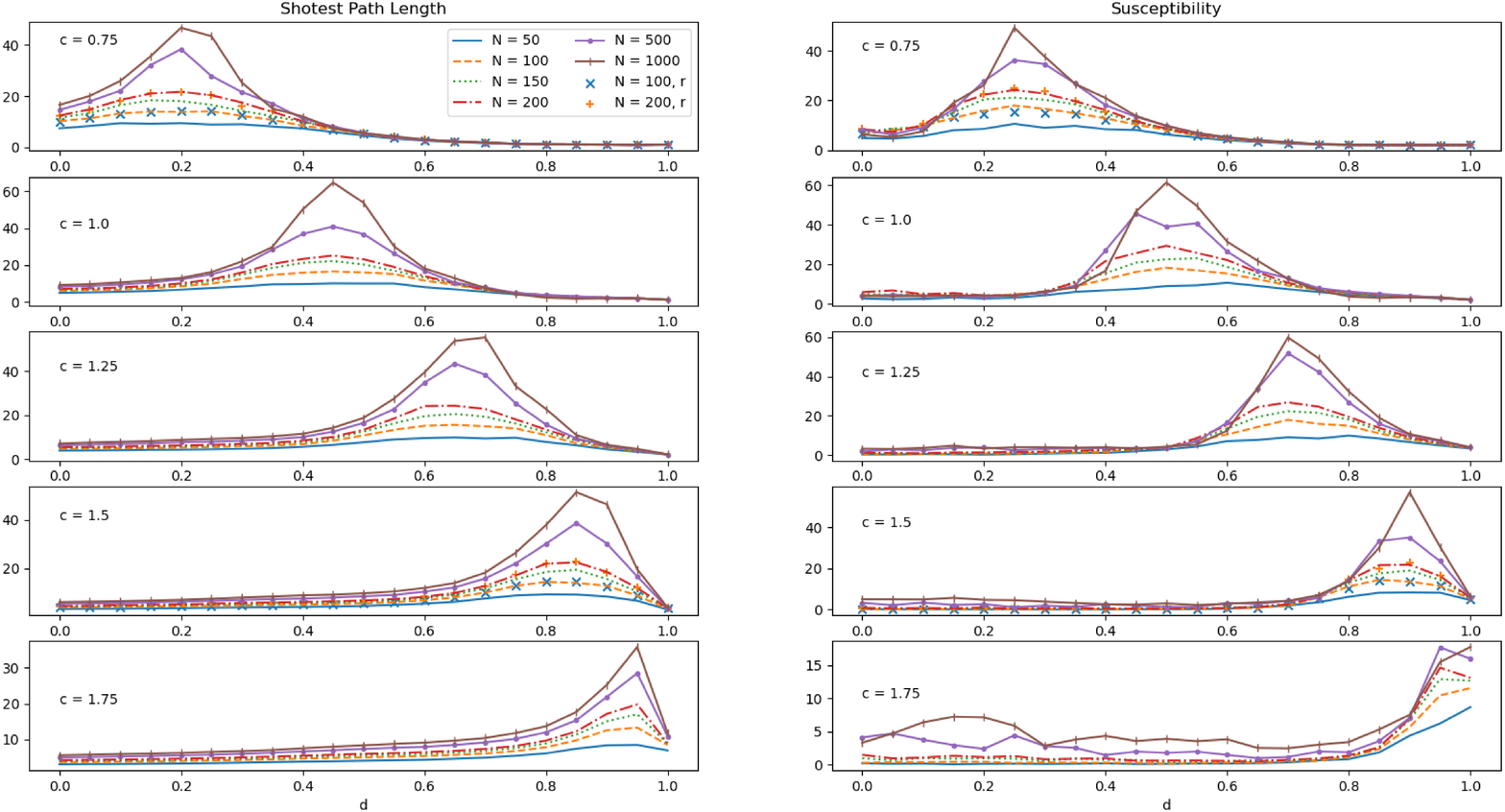}
\caption{The shortest path length and the susceptibility as functions of the probabilities $d$ and $(1-d)$ of the dictator game and ultimatum game being played by the agents, the number of which is $N=$ 50, 100, 150, 200, 500, and 1000. Here the living cost parameter $c=$ 0.75, 1.0, 1.25, 1.5, and 1.75. The results of tests with random ordering are also shown here, with $c = 0.75, 1.5$ and $N = 100, 200$, marked with $r$ in the legend. } 
\label{udfig}
\end{center}
\end{figure*}

In Fig. \ref{netFig1} we present the simulation results for the structure of the co-evolving social network of $N=100$ agents having played the dictator and ultimatum game with equal probability $d = 0.5$ and for three different values of the cost parameter $c=$ 0.3, 1.1, and 3.0 for the total of $10000$ time steps. Here one observes that when the cost parameter $c$ is increased the system experiences a structural change from a collection of mostly simple pairs of agents to longer chains and eventually to
connected complex networks. Similar structural phase change or transition is observed for all values of the probability $d$. In essence what happens is that the short chains of agents lengthen until they start joining to other chains and become more and more entangled and then networked. This resembles the phase transition due to the crystallization process in freezing liquids.

In order to investigate such a network formation process in more detail we focus our attention to the shortest path length $L$ and to the susceptibility $\langle s \rangle$, of which the latter is defined as the second moment of $s$-sized clusters, excluding the largest connected component of the network:
\begin{equation}
\langle s \rangle = \frac{\sum_s n_s s^2 }{\sum_s n_s s}.
\label{suscep}
\end{equation}
The susceptibility illustrates fragmentation of the social network, measuring as it does the size of subsidiary networks. To calculate $L$ and $\langle s \rangle$, we have done the simulations up to 10000 time steps, letting the system to reach the stationary state for the first 5000 time steps and then taking the averages over the last 5000 time steps for different numbers of agents $N =$ $50$, $100$, $150$, $200$, $500$ and  $1000$. For sufficiently reliable statistics we have performed $100$ simulation runs for agent populations up to $N = 200$, while for $N =500, 1000$ we could reasonably perform only $10$ and $5$ runs, respectively. It should be noted that in systems showing phase transition near their phase transition or critical point, there are large fluctuations and the time evolution of the system tend to show critical slowing-down phenomenon. In this situation to get accurate results one would need to run the system longer, which is unfortunately beyond the computational resources at our disposal. Nevertheless, we believe that our calculations of $L$ and $\langle s \rangle$ are sufficiently accurate in characterising the social structure and network formation process. 

In Fig. \ref{phasefig} we show the averages of the shortest path length and susceptibility for the pure dictator ($d = 1$) and ultimatum game ($d = 0$) models as functions of the cost $c$ for several agent populations $N$. For the ultimatum game strategy (Fig. \ref{phasefig} a) and b)) we observe a strong peak in both $L$ and $\langle s \rangle$ at $c \approx 0.6$, which becomes stronger with increasing number $N$ agents. The fact that the height of the peak increases with the size of the network suggests that in the network formation process the change in structure could be characterised as a phase transition. The situation is quite similar for the dictator game strategy (Fig. \ref{phasefig} c) and d)) but now the peaks of $L$ and $\langle s \rangle$ are broader such that the point of phase change or transition is less well defined, occurring at  $c \approx 2$.

Next we investigate the behaviour of the hybrid model in situation where the agents play with different probabilities or proportions of the ultimate and dictator game strategies. In Fig. \ref{udfig} we depict the shortest path length and susceptibility as a function of the probability $d$ for five different values of the cost parameter $c=$ 0.75 1.0, 1.25, 1.5, and 1.75. In this figure we observe broad peaks in both quantities for all the cost parameter $c$ values. As observed in Fig. \ref{phasefig} as the function of $c$, both the shortest path length $L$ and susceptibility $\langle s \rangle$ peak rises strongly when the number of agents $N$ in the network increases. Also a significant feature is that the peak seems to occur at higher values of probability $d$ as the cost $c$ is increased, such that for $c = 0.75$ the peak occurs at $d \approx 0.25$ while for $c = 1.75$ at $d \approx 0.95$. 

In Fig. \ref{udfig} we also show the results of tests we performed to study, whether having a random or fixed playing order in our model makes any difference in the results. We chose two values of $N$ ($100$ and $200$) and $c$ ($0.75$ and $1.5$), respectively, and made a set of four runs matching the standard fixed order runs, in which the agents played the game in random order instead of fixed order. As can be seen, the two playing orders produce essentially identical results, with the greatest deviations between the two being found near the phase transition or critical points. This suggests that our assumption of the relative irrelevance of the playing order is basically correct, since greater variability, even with $100$ realisations, is to be expected near the critical point.

\begin{figure}[ht]
\begin{center}
\epsfxsize = 1.0\columnwidth \epsfysize = 0.5\textheight \epsffile{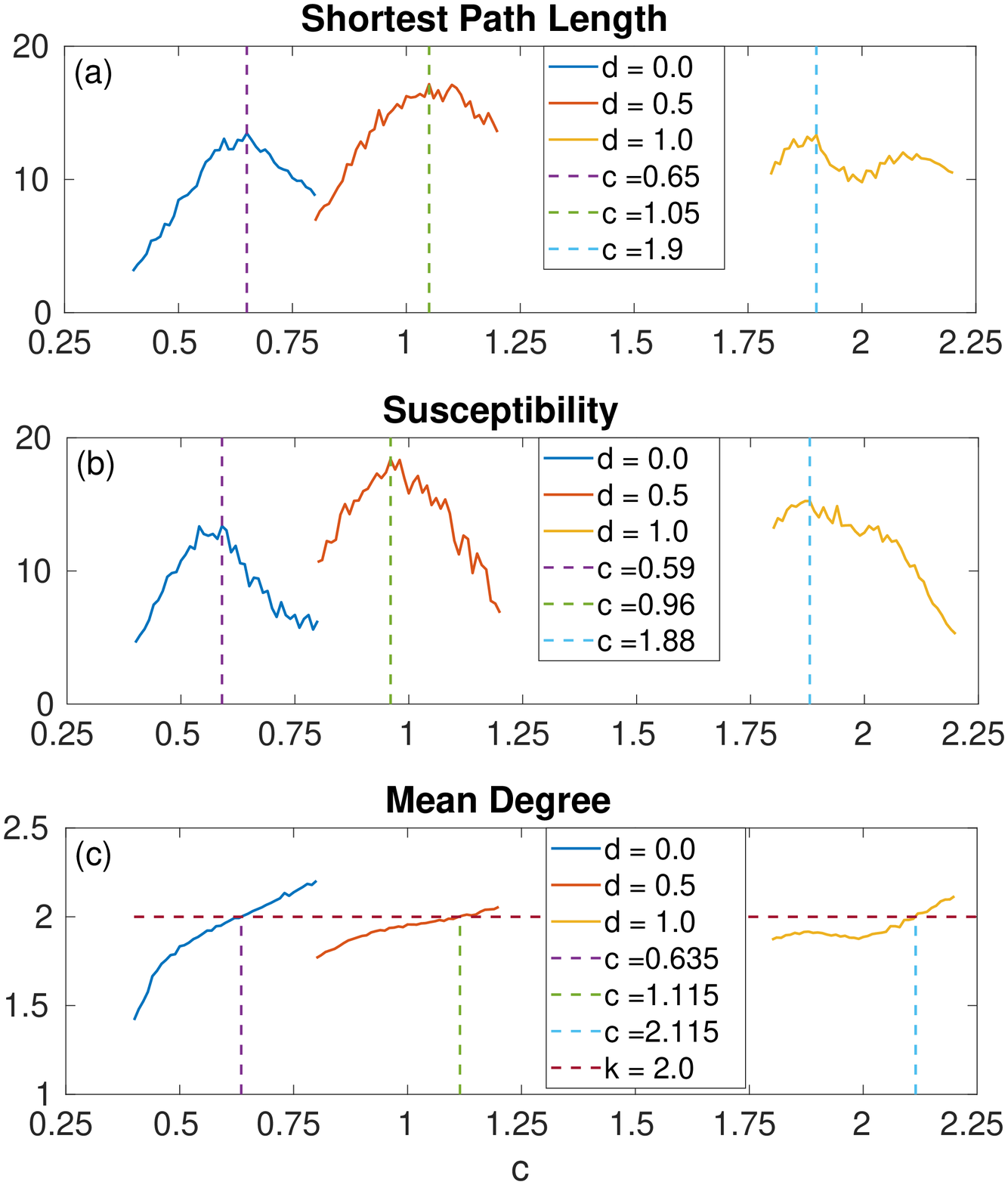}
\caption{The shortest path length (panel (a)), susceptibility (panel (b)) and average degree (panel (c)) near the phase transition points for pure ultimatum ($d = 0$), equal mix of ultimatum and dictator ($d=0.5$) and pure dictator game strategies ($d=1$). The vertical dashed lines show either maximal values of the shortest path length and susceptibility, or the points at which the average degree exceeds the value of $2$.}
\label{crit}
\end{center}
\end{figure}

In Fig. \ref{crit} we take a closer look at how the average degree, the shortest path length and the susceptibility behave for three different game strategy proportions, i.e. $d = 0$ (pure ultimatum strategy), $d=0.5$ (equal mix of ultimatum and dictator strategies) and $d=1$ (pure dictator strategy) in the neighbourhood of the respective phase transition points, where the shortest path length and susceptibility are reaching their maximal values, marked in Fig. \ref{crit} (a) and (b) with vertical dashed lines. In Fig. \ref{crit} (c) we approximate the phase transition points (marked also with vertical lines) to occur, when the average degree rises above the value of $2$, because up to that point the social structure is mostly chain-like and after that it becomes more entangled and network, i.e. network formation takes place. We see that for the pure ultimatum strategy ($d = 0$) case the maxima of the shortest path length and susceptibility occur relatively close to the value of the cost $c$ at which the average degree rises above $2$, while for equal mix of ultimatum and dictator strategy ($d = 0.5$) and for pure dictator strategy ($d = 1$) the values of the cost $c$ seem to deviate of the order of 15 \%. 

\begin{figure}[ht]
\begin{center}
\epsfxsize = 1.0\columnwidth \epsfysize = 0.25\textheight \epsffile{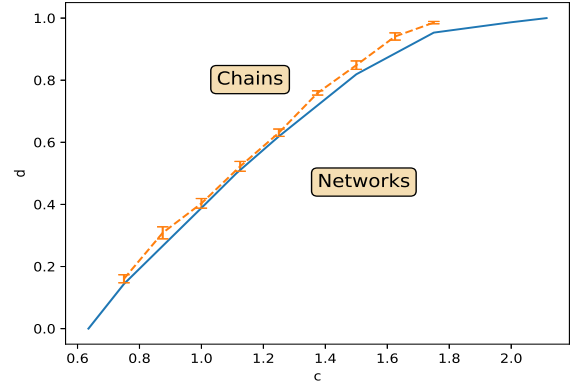}
\caption{Phase diagram in the $(c,d)$-plane for N = 100, calculated by using two different methods. The solid line shows the estimate from linear interpolation, while the dashed line with the error bars shows the results of the Monte Carlo method.}
\label{diagram}
\end{center}
\end{figure}

The results presented in Figs. \ref{phasefig}, \ref{udfig} and \ref{crit} allow us to draw a rough phase diagram for the system, as depicted in Fig. \ref{diagram}, for the pair of model parameters, i.e. the living cost parameter $c$ and relative proportions, $d$ and $(1-d)$ of the dictator and ultimatum games, respectively. The phases depicted in the figure concern the changes in the network structure such that in the upper part of the curve one obtains mostly chains, while in the lower part there are networks of agents connected with more than two neighbors. The point at which the average degree of agents surpasses the value $2$ is taken to represent the onset of the complex network formation. Since the exact parameter values at which this happens are essentially never captured in the numerical data, the linear interpolation is used to estimate where the crossings of average degree value $2$ take place. The resultant curve, depicted as a solid line in Fig. \ref{diagram}, runs approximately from point $(0.6,0)$ to $(2.1,1)$, in a monotonically rising but sub-linear manner. 

Due to the rough nature of the linear interpolation used to construct the phase transition (solid) line in Fig. \ref{diagram}, we were not able to make any reasonable error estimates in this case. To gain some idea on the size of the errors, we used a Monte Carlo method to obtain an alternative  phase transition curve and its error estimates. We did this by selecting nine values of the cost parameter, listed in Table \ref{f5tab}, and then by randomly choosing values for $d$ in the close vicinity of the phase curve indicated by the linear interpolation method at these points. Running the simulations with $N = 100$ agents for $100$ times we got a value for the average degree for thus chosen pair of parameters, and if this exceeded the value of $2$ by small enough margin (less than $0.1$ to be exact), the parameter pair was stored. Repeating these steps we obtained for each of the chosen values of $c$ an ensemble of values of $d$, for which the mean and standard deviation could be calculated. If the ensemble seemed to terminate at the limits of the search area, the search area was expanded until the ensemble fitted within it. The mean values for $d$ acquired in this way are illustrated in Fig. \ref{diagram} with a dashed line, and the standard deviations provide the error bar estimates. These results are also summarised in Table \ref{f5tab} along with the ensemble sizes they are based upon. 

As seen in Fig. \ref{diagram}, the results of the Monte Carlo method agree rather closely with those obtained by linear interpolation, though the difference between the methods, small as it is, increases as $c$ increases. The errors, as given by the Monte Carlo method turn out to be relatively small: at their highest, they only amount to little more than $8 \%$. The convergence of the results of the two methods and the small errors mean that the phase diagram shown in Fig. \ref{diagram} can be considered reasonably accurate. 

The reason why the complex network structure formation occurs at different points for the ultimatum and dictator game models is that in the former case, the agents need to take into account the possibility of rejection when making their proposals, while in the latter case they do not need to do that. For any values of the relative proportions $d$ and $(1-d)$ of the dictator and ultimatum games being played, respectively, the network formation occurs when the number of neighbours needed to make profit is more than $2$, and so the chains change into complex social networks. 
In the case of the dictator game an agent playing with others with similar offering rates can expect to see a profit of $M$ per playing partner. Thus it needs at least $\lfloor c \rfloor$ neighbours to offset the living costs (as argued in \cite{SGKBK2018}). Consequently it is there where the complex network structure formation occurs, as depicted in Fig. \ref{diagram}. When $d < 1$, the agents need more than $\lfloor c \rfloor$ relations because some of them might decide to turn down the offers and consequently the network structure formation will occur at a lower value of the living cost parameter $c$. 

\begin{table}[]
    \centering
    \begin{tabular}{c|c|c|c|c}
          c  &  F & mean(d) & std(d) & \% \\
          \hline
         0.75  & 101 & 0.161 & 0.013 & 8.052 \\
         0.875 & 60  & 0.309 & 0.019 & 6.288 \\
         1.0   & 116 & 0.403 & 0.016 & 3.863 \\
         1.125 & 115 & 0.523 & 0.016 & 3.027 \\
         1.25  & 115 & 0.631 & 0.012 & 1.874 \\
         1.375 & 88  & 0.759 & 0.007 & 0.916 \\
         1.5   & 104 & 0.849 & 0.014 & 1.606 \\
         1.625 & 109 & 0.941 & 0.012 & 1.243 \\
         1.75  & 48  & 0.986 & 0.004 & 0.376 \\
    \end{tabular}
    \caption{The data obtained from the Monte Carlo simulations for Fig. \ref{diagram}. Listed here are the fixed values of $c$ chosen, the number of acceptable values of $d$ found $(F)$, along the means and standard deviations of these values. Also listed is the percentual error $(\%)$. } 
    \label{f5tab}
\end{table}

\section{Conclusions}
\label{con}

In this work we have studied the effects of internalised norms to social structure formation in an agent based setting, and found that more reciprocal norms aid community formation. The agent-based model we created involved agents playing a hybrid of dictator and ultimatum games, which represent unilateralist and limitedly reciprocal social strategies, respectively. The social motivations of the agents revolve around their tendency to maximize their socioeconomic standing, and they redefine mutual social relations based on their individual utilities.

Our main focus has been on the observed structural change in the social connectivity of agents taking place from the chains to more complex network structure when the living cost parameter $c$ is increased. This structural change is interpreted as a phase transition as it manifests itself as peaks in the shortest path length and susceptibility of the co-evolved social network. We have found that for both the pure ultimatum game and dictator game strategies, this transition occurs for the cost of living of $c \approx 0.6$ and $c \approx 2$, respectively. Setting the parameter $c$ between these values and plotting the susceptibility and shortest path length as a function of the relative proportions $d$ and ($1-d$) of the dictator and ultimatum game strategies, respectively, revealed a broad peak that indicates a similar network formation taking place in this case as well.

The discovery of the shift in the transition point of social structure formation as a function of the relative proportions, $d$ and ($1-d$), the two strategies have been played has implications for the applicability of BTH-style models to real human societies, the most profound of which relates to the role of the ground rules of social interactions in the formation of social structures. The fact that the threshold for complex network structure formation has a lower value for the living cost parameter $c$ when the ultimatum game becomes more prevalent, means that in that case the agents are generally more inclined towards forming complex societies than when the dictator game is predominant. As the ultimatum game could be characterised to be more reciprocal than the dictator game, given that both players get to make a decision regarding the division of funds, one could draw the conclusion that reciprocity in social interactions makes it easier for the agents to form more complex communities. This is, indeed, a promising result for BTH, since one would intuitively expect this to be the case in the real human societies. Finally judging from our numerical results, the simulated societies based on purely dictatorial social interactions need more than three times the amount of external pressure to maintain complex social structures than those based on ultimatum type interactions.

In order to further study the effect of reciprocity on social structure formation under BTH one would need to define a measure for the reciprocity of social interaction types, such as the dictator and ultimatum games, and then study the propensity of social structure formation as a function of this measure. However, defining such a measure is beyond the scope of the present study, as is the devising of new social interactions that would have different ratings on that measure, as would be necessary for the aforementioned future research. 

\acknowledgements

JES acknowledges the financial support of Niilo Helander's foundation. The computational resources provided by the Aalto Science-IT project have been utilised in this work. KK acknowledges support from the Rutherford Foundation Visiting Fellowship at The Alan Turing Institute, UK, and from the European Community’s H2020 Program under the scheme INFRAIA-1-2014-2015: Research Infrastructures”, Grant agreement No. 654024 SoBigData: Social Mining and Big Data Ecosystem” (http://www. sobigdata.eu). RAB wants to acknowledge financial support from Conacyt (Mexico) through project 283279. We thank J\'{a}nos Kert\'{e}sz for his suggestions at the early stage of this work.

\bibliographystyle{unsrt}
\bibliography{BTHgeneral}

\begin{thebibliography}{10}

\bibitem{AHH2015}
C.~Anderson, J.~A.~D. Hildreth, and L.~Howland.
\newblock Is the desire for status a fundamental human motive? a review of the
  empirical literature.
\newblock {\em Psychological Bulletin}, 141(3):574--601, 2015.

\bibitem{ZTCB2008}
C.~F. Zink, Y.~Tong, Q.~Chen, D.~S. Bassett, J.~L. Stein, and
  A.~Meyer-Lindenberg.
\newblock Know your place: Neural processing of social hierarchy in humans.
\newblock {\em Neuron}, 58(2):273--283, 2008.

\bibitem{ISS2008}
K.~Izuma, D.~N. Saito, and N.~Sadato.
\newblock Processing of social and monetary rewards in the human striatum.
\newblock {\em Neuron}, 58(2):284--294, 2008.

\bibitem{KMD2012}
Dharshan Kumaran, Hans~Ludwig Melo, and Emrah Duzel.
\newblock The emergence and representation of knowledge about social and
  nonsocial hierarchies.
\newblock {\em Neuron}, 76(3):653 -- 666, 2012.

\bibitem{UP2014}
Amanda~V. Utevsky and Michael~L. Platt.
\newblock Status and the brain.
\newblock {\em PLOS Biology}, 12(9):1--4, 09 2014.

\bibitem{ASGB2014}
Andreas Angourakis, José~Ignacio Santos, José~Manuel Galán, and Andrea~L.
  Balbo.
\newblock Food for all: An agent-based model to explore the emergence and
  implications of cooperation for food storage.
\newblock {\em Environmental Archaeology}, 20(4):349--363, 2015.

\bibitem{SGGBK2017}
Jan~E Snellman, Gerardo I{\~n}iguez, Tzipe Govezensky, R~A Barrio, and Kimmo~K
  Kaski.
\newblock Modelling community formation driven by the status of individual in a
  society.
\newblock {\em Journal of Complex Networks}, 5(6):817--838, 2017.

\bibitem{adler}
A.~Adler.
\newblock {\em The Practice and Theory of Individual Psychology}.
\newblock Routledge, Trench and Trubner \& Co, Ltd, 1924.
\newblock Reprint. Abingdon: Routledge (1999).

\bibitem{SGKBK2018}
Jan~E Snellman, Gerardo I{\~n}iguez, J{\'a}nos Kert{\'e}sz, Rafael~A Barrio,
  and Kimmo~K Kaski.
\newblock Status maximization as a source of fairness in a networked dictator
  game.
\newblock {\em Journal of Complex Networks}, page cny022, 2018.

\bibitem{HN2006}
Petter Holme and M.~E.~J. Newman.
\newblock Nonequilibrium phase transition in the coevolution of networks and
  opinions.
\newblock {\em Phys. Rev. E}, 74:056108, Nov 2006.

\bibitem{IKKB2009}
G.~I{\~n}iguez, J.~Kert{\'e}sz, K.~K. Kaski, and R.~A. \&~Barrio.
\newblock Opinion and community formation in coevolving networks.
\newblock {\em Physical Review E}, 80:066119, 2009.

\bibitem{LGAGI2005}
M.F. Laguna, S.~Risau Gusman, G.~Abramson, S.~Gonçalves, and J.R. Iglesias.
\newblock The dynamics of opinion in hierarchical organizations.
\newblock {\em Physica A: Statistical Mechanics and its Applications},
  351(2):580 -- 592, 2005.

\bibitem{JA2005}
Wander Jager and Fr{\'e}d{\'e}ric Amblard.
\newblock Uniformity, bipolarization and pluriformity captured as generic
  stylized behavior with an agent-based simulation model of attitude change.
\newblock {\em Computational {\&} Mathematical Organization Theory},
  10(4):295--303, Jan 2005.

\bibitem{GK2006}
A.~Grabowski and R.A. Kosiński.
\newblock Ising-based model of opinion formation in a complex network of
  interpersonal interactions.
\newblock {\em Physica A: Statistical Mechanics and its Applications},
  361(2):651 -- 664, 2006.

\bibitem{SM2013}
Zhanli Sun and Daniel Müller.
\newblock A framework for modeling payments for ecosystem services with
  agent-based models, bayesian belief networks and opinion dynamics models.
\newblock {\em Environmental Modelling \& Software}, 45:15 -- 28, 2013.
\newblock Thematic Issue on Spatial Agent-Based Models for Socio-Ecological
  Systems.

\bibitem{IGDKB2014}
G.~I{\~n}iguez, T.~Govezensky, R.~Dunbar, K.~Kaski, and R.~A Barrio.
\newblock Effects of deception in social networks.
\newblock {\em Proceedings. Biological sciences}, 281:20141195, 2014.

\bibitem{BGDIK2015}
R.~A. Barrio, T.~Govezensky, R.~Dunbar, G.~I{\~n}iguez, and K~Kaski.
\newblock Dynamics of deceptive interactions in social networks.
\newblock {\em Journal of the Royal Society, Interface}, 112:20150798, 2015.

\bibitem{EZ2000}
V\'{\i}ctor~M. Egu\'{\i}luz and Mart\'{\i}n~G. Zimmermann.
\newblock Transmission of information and herd behavior: An application to
  financial markets.
\newblock {\em Phys. Rev. Lett.}, 85:5659--5662, Dec 2000.

\bibitem{W2015}
Wilson Perez-Oviedo.
\newblock Citizens, dictators and networks: A game theory approach.
\newblock {\em Rationality and Society}, 27(1):3--39, 2015.

\bibitem{RB2002}
R\'eka Albert and Albert-L\'aszl\'o Barab\'asi.
\newblock Statistical mechanics of complex networks.
\newblock {\em Rev. Mod. Phys.}, 74:47--97, Jan 2002.

\bibitem{DM2002}
S.~N. Dorogovtsev and J.~F.~F. Mendes.
\newblock Evolution of networks.
\newblock {\em Advances in Physics}, 51(4):1079--1187, 2002.

\bibitem{N2003}
M.~Newman.
\newblock The structure and function of complex networks.
\newblock {\em SIAM Review}, 45(2):167--256, 2003.

\bibitem{SPS2012a}
Attila Szolnoki, Matjaž Perc, and György Szabó.
\newblock Accuracy in strategy imitations promotes the evolution of fairness in
  the spatial ultimatum game.
\newblock {\em EPL (Europhysics Letters)}, 100(2):28005, 2012.

\bibitem{SPS2012b}
A.~{Szolnoki}, M.~{Perc}, and G.~{Szab{\'o}}.
\newblock {Defense Mechanisms of Empathetic Players in the Spatial Ultimatum
  Game}.
\newblock {\em Physical Review Letters}, 109(7):078701, August 2012.

\bibitem{SP2013}
Attila Szolnoki and Matja\ifmmode \check{z}\else~\v{z}\fi{} Perc.
\newblock Correlation of positive and negative reciprocity fails to confer an
  evolutionary advantage: Phase transitions to elementary strategies.
\newblock {\em Phys. Rev. X}, 3:041021, Nov 2013.

\bibitem{EMV2006}
George C. M.~A. Ehrhardt, Matteo Marsili, and Fernando Vega-Redondo.
\newblock Phenomenological models of socioeconomic network dynamics.
\newblock {\em Phys. Rev. E}, 74:036106, Sep 2006.

\bibitem{MJTKK2019}
Yohsuke Murase, Hang-Hyun Jo, J{\'a}nos T{\"o}r{\"o}k, J{\'a}nos Kert{\'e}sz,
  and Kimmo~K Kaski.
\newblock Structural transition in social networks: The role of homophily.
\newblock {\em Scientific Reports}, 9(1):2045--2322, 2019.

\bibitem{CCV2016}
Nestor {Caticha}, Rafael {Calsaverini}, and Renato {Vicente}.
\newblock {Phase transition from egalitarian to hierarchical societies driven
  by competition between cognitive and social constraints}.
\newblock {\em arXiv e-prints}, page arXiv:1608.03637, Aug 2016.

\bibitem{KKT1986}
D.~Kahneman, J.~L. Knetsch, and R.~H. Thaler.
\newblock Fairness and the assumptions of economics.
\newblock {\em The Journal of Business}, 59(4):285--300, 1986.

\bibitem{GT1990}
W.~G\"uth and R.~Tietze.
\newblock Ultimatum bargaining behavior: A survey and comparison of
  experimental results.
\newblock {\em Jornal of Econonomic Psychology}, 11(3):417--449, 1990.

\bibitem{S1955}
Herbert~A. Simon.
\newblock A behavioral model of rational choice.
\newblock {\em The Quarterly Journal of Economics}, 69(1):99--118, 1955.

\bibitem{HS}
Herbert Simon.
\newblock Bounded rationality and organizational learning.
\newblock {\em Organization Science}, 2(1):125--134, 1991.

\bibitem{E2011}
Christoph Engel.
\newblock Dictator games: a meta study.
\newblock {\em Experimental Economics}, 14(4):583--610, Nov 2011.

\bibitem{metalyysi}
H.~Oosterbeek, R.~Sloof, and G.~van~de Kuilen.
\newblock Cultural differences in ultimatum game experiments: Evidence from a
  meta-analysis.
\newblock {\em Experimental Economics}, 7(2):171--188, 2004.

\bibitem{CC2008}
Juan~Camilo Cardenas and Jeffrey Carpenter.
\newblock Behavioural development economics: Lessons from field labs in the
  developing world.
\newblock {\em The Journal of Development Studies}, 44(3):311--338, 2008.

\bibitem{FLK2015}
Feng Chunliang, Luo Yue‐Jia, and Krueger Frank.
\newblock Neural signatures of fairness‐related normative decision making in
  the ultimatum game: A coordinate‐based meta‐analysis.
\newblock {\em Human Brain Mapping}, 36(2):591--602.

\bibitem{X2010}
B.~Xianyu.
\newblock Social preference, incomplete information, and the evolution of
  ultimatum game in the small world networks: An agent-based approach.
\newblock {\em Journal of Artificial Societies and Social Simulation}, 13(2):7,
  2010.

\end{thebibliography}

\end{document}